%
%
%
%
%
%
%
%
%
%
%
%
%
%
\input phyzzx
%
%
\catcode`\@=11 
\def\papersize{\hsize=40pc \vsize=53pc \hoffset=0pc \voffset=-2pc
   \advance\hoffset by\HOFFSET \advance\voffset by\VOFFSET
   \pagebottomfiller=0pc
   \skip\footins=\bigskipamount \normalspace }
\catcode`\@=12 
\papers
\vsize=23.cm
\hsize=15.cm

\Pubnum={LPTENS-97/63 \cr
{\tt hep-th@xxx/9801047} \cr
January 1998}

\date={}
\pubtype={}
\titlepage
\title{{\bf A comment on compactification of M-theory\break 
on an (almost) light-like circle}
}
\author{Adel~Bilal
\foot{Partially supported by the
European Commision under TMR contract FMRX-CT96-0090.}
}
\vskip .5cm
\address{
CNRS - Laboratoire de Physique Th\'eorique de l'\'Ecole
Normale Sup\'erieure
\foot{{\rm unit\'e propre du CNRS, associ\'ee \`a l'\'Ecole Normale
Sup\'erieure et l'Universit\'e Paris-Sud}}    \break
24 rue Lhomond, 75231 Paris Cedex 05, France  \break
{\tt bilal@physique.ens.fr}
}

\vskip 0.5cm
\abstract{In perturbative quantum field theory the limit of compactification on 
an almost light-like circle has recently been shown to be plagued by 
divergences. We argue that the light-like limit for M-theory probably 
is free of such divergences due to, among others, the existence of 
the wrapping modes of the membranes. To illustrate this, we consider 
superstring theory compactified on an almost light-like circle. \nextline
\indent 
Specifically, we compute a one-loop four-point amplitude in type II theory. 
As is well known, if the external states have {\it vanishing} momenta 
in the compact dimension, the divergence in the light-like limit is even 
stronger than in field theory. However, in the case of present interest, 
where these external momenta are {\it non-vanishing}, there is a subtle 
compensation and the resulting amplitude has a well-defined and 
finite light-like limit. The net effect of taking the light-like limit is to 
replace the integration over one of the moduli of the 4-punctured torus by 
a sum over a discrete modulus taking values in a finite lattice on the torus. 
The same result can also be obtained from a suitably ``Wick rotated" 
amplitude computed directly with a compact light-like circle.
}

%

\endpage
\pagenumber=1

\def\PL #1 #2 #3 {Phys.~Lett.~{\bf #1} (#2) #3}
\def\NP #1 #2 #3 {Nucl.~Phys.~{\bf #1} (#2) #3}
\def\PR #1 #2 #3 {Phys.~Rev.~{\bf #1} (#2) #3}
\def\PRL #1 #2 #3 {Phys.~Rev.~Lett.~{\bf #1} (#2) #3}
\def\CMP #1 #2 #3 {Comm.~Math.~Phys.~{\bf #1} (#2) #3}

\def\b{\beta}

\def\m{\mu}
\def\n{\nu}
\def\nsr{\nu_{sr}}
\def\r{\rho}

\def\rmd{{\rm d}}
\def\rd{\sqrt{2}}

\def\tr{{\rm tr}\, }

\def\pe{p_{11}}
\def\re{R_{11}}
\def\xe{x^{11}}
\def\rs{R_s}
\def\e{\epsilon}
\def\xm{x^-}
\def\xp{x^+}

\def\t{\tau}
\def\tb{{\overline\tau}}
\def\ap{\alpha'}
\def\rap{\sqrt{\ap}}
\def\rmd{{\rm d}}
\def\to{\rightarrow}
\def\Im{{\rm Im}}
\def\Re{{\rm Re}}
\def\kcl{K_{\rm cl}}

\REF\WIT{E. Witten, {\it String theory dynamics in various
dimensions}, \NP B443 1995 85 , {\tt hep-th/9503124}.}

\REF\BFSS{T. Banks, W. Fischler, S.H. Shenker and L. Susskind, {\it
M theory as a matrix model: a conjecture}, \PR D55 1997 5112 , 
{\tt hep-th/9610043}.}

\REF\AB{A. Bilal, {\it M(atrix) theory : a pedagogical introduction}, {\tt hep-th/9710136}.}

\REF\CJ{E. Cremmer, B. Julia and J. Scherk,  {\it Supergravity theory in 11 dimensions}, \break
\PL 76B 1978 409 .}

\REF\SUSS{L. Susskind,  {\it Another conjecture about M(atrix) theory}, {\tt hep-th/9704080}.}

\REF\SEI{N. Seiberg, {\it Why is the matrix model correct?}, {\tt hep-th/9710009}.}

\REF\SEN{A. Sen,  {\it D0-branes on $T^n$ and matrix theory}, {\tt hep-th/9709220}.}

\REF\PH{S. Hellerman and J. Polchinski, {\it Compactification in the lightlike limit},  {\tt hep-th/9711037}.}

\REF\GSW{M. Green, J. Schwarz and E. Witten, {\it Superstring theory}, vol. 2, Cambridge University
Press, 1987.}

{\bf \chapter{Introduction and summary}}

\section{Introduction}

M-theory compactified on a circle of 
radius $\re$ is  type IIA superstring theory with coupling $g_s=\re/\sqrt{\ap}$ [\WIT]. 
Banks, Fischler, Shenker and Susskind  [\BFSS] conjectured that when the
theory is boosted to the infinite momentum frame in the $\xe$ direction, the only relevant degrees of
freedom are D0-branes, and M-theory then is described by a ten-dimensional ${\rm U}(N)$ 
super YM theory reduced to $0+1$ dimensions, i.e. matrix quantum mechanics (for a pedagogical
review, see ref. \AB). 
The momentum in the
$\xe$ direction is $\pe={N\over \re}$ so that the infinite momentum limit is obtained by letting
$N\to\infty$. The low-energy sector for large $\re$ should then describe eleven-dimensional
supergravity [\CJ].

A little later, Susskind [\SUSS] suggested to consider M-theory compactified on a light-like circle,
identifying $\xm\equiv (x^0-\xe)/\rd$ with $\xm +2\pi R$. One should keep in mind that the proper length of
this circle vanishes and that the parameter $R$ has no invariant meaning since it can be changed by a
Lorentz boost. The momenta $p_-={N\over R}$ are discrete and the conjecture states that the
discrete light-cone quantization (DLCQ) of M-theory in a sector of fixed total $p_-$ 
is again given by a ${\rm U}(N)$ matrix quantum
mechanics as obtained by reduction from the ten-dimensional super YM theory. This formalism has
the advantage that various dualities are already manifest at finite $N$.

In an insightful paper, Seiberg [\SEI] (see also Sen [\SEN]) has related both approaches to M-theory by
considering the light-like compactification on a (null) circle with radius $R_0$ as being
obtained in the limit of a 
very large boost from a space-like compactification on a circle of very small radius $\rs=\re$. More
precisely, a space-like circle
$\xe \simeq \xe + 2\pi \rs$
%
when subject to a very large boost 
becomes an almost light-like circle
$\xm\simeq \xm + 2\pi R_0 \ , \  \xp \simeq \xp + \pi R_s^2/R_0$.
%
In the limit $\rs\to 0$ with $R_0$ fixed, the boost becomes infinite and
the latter circle is really light-like, while the space-like circle in the $\xe$ direction has
shrunk to zero length. Using this very large boost combined with scaling arguments, Seiberg argued
that the DLCQ of M-theory should be interpreted as being equivalent to M-theory compactified on a very
small space-like circle. The latter is the IIA string at  weak coupling and in a sector of non-zero D0-brane
charge where only the open string ground states survive while the oscillators decouple. This must indeed
be described by the matrix model.

\section{Motivation}

As appealing as this argument is, one might feel uneasy about infinite boosts, or else about
approximating  a light-like circle by an almost light-like, i.e. still space-like circle. To elucidate further
whether one might trust such an approximation, Hellerman and Polchinski  [\PH] have studied some
loop diagrams in quantum {\it field} theory when compactified on an almost light-like circle. To do so,
they introduce a parameter $\e$ in the (flat) space-time metric such that for $\e\to 0$ the circle is
truely light-like ($\e \approx R_s/R_0$ and $2\pi\e R_0$ is the proper length of the compact direction).
Specifically, start in ten-dimensional Minkowski space with metric 
$\rmd s^2=- (\rmd  x^0)^2 + (\rmd  x^1)^2 + (\rmd  x^i)^2 $ ($i=2,\ldots 9$) and
compactify $ x^1$ on a circle of radius $\e R_0$:
$$  x^1 \simeq   x^1 + 2\pi \e R_0 \quad , \quad   x^0 \simeq  x^0 \ .
\eqn\i$$
(The ``transverse" $x^i$ are unaffected in all what follows.) Next, one makes a large boost with 
parameter $\b = (1-\e^2/2)/(1+\e^2/2)$ so that the boosted coordinates  $\tilde x^0$ and $\tilde x^1$ satisfy
$\tilde x^1 \simeq \tilde x^1 -(1+\e^2/2) 2\pi R_0 /\rd \quad , 
\quad \tilde x^0 \simeq \tilde x^0 + (1-\e^2/2) 2\pi R_0 /\rd $, 
the metric being unaffected. Introduce $x^\pm=(\tilde x^0\pm \tilde x^1)/\rd$. Then
$$ \xm\simeq \xm +2\pi R_0 \quad , \quad \xp\simeq\xp - \e^2 \pi R_0 
\quad , \quad \rmd s^2=-2 \rmd\xp\rmd\xm + (\rmd x^i)^2\ .
\eqn\ii$$
As in [\PH]  introduce $t=\xp +\e^2 \xm/2$ so that
$$ \xm\simeq \xm +2\pi R \quad , \quad t\simeq t 
\quad , \quad \rmd s^2=-2 \rmd t\rmd\xm + \e^2 \rmd\xm\rmd\xm + (\rmd x^i)^2\ .
\eqn\iii$$
Note that  $g^{t-}=g^{-t}=-1,\  g^{tt}=-\e^2, \  g^{--}=0$, so that $p^t=-p_--\e^2
p_t$.
Equations \ii\ and \iii\ then clearly show that in the $\e\to 0$ limit, 
the circle becomes truely light-like with radius $R_0$,
while the equivalent space-like circle \i\ has shrunk to zero size. Note that  $t=\e  x^0$ and
$\xm= ( x^0- x^1)/\e$. Obviously, the momentum $p_-=n/R_0$ is discretised, while $p_t$ is not.
How do the momenta in the compact space-like circle ($p_0$)  and compact light-like circle
($p_-$) compare?
From $p_0\, x^0+p_1\,  x^1=p_-\,  x^-+p_t\,  x^t={n\over R_0} {x^0-x^1\over \e} + p_t\,  \e x^0$ we get
$$p_0=\e\,  p_t +{n\over \e R_0} \quad , \quad -p_1 ={n\over \e R_0} \ .
\eqn\iv$$
The point we want to stress is that the integer $n$ that characterizes the compact light-like momentum
$p_-$ is the {\it same} integer\foot{up to a sign flip which we could have avoided by defining $x^-$ and
hence $p_-$ with the opposite sign}
as the $n$ that characterizes the space-like momentum $p_1$. So for fixed
light-like momentum $p_-$, we have a fixed $n$ and hence the momentum $p_1$ in the compact space-like 
dimension must be taken to diverge as $\e\to 0$. This will be important below.

Hellerman and Polchinski find [\PH] that in a generic
$D$-dimensional QFT, loop diagrams with vanishing $p_-$ exchange 
diverge as ${1\over \e}$ when $\e\to 0$. This is
due to the longitudinal zero-modes becoming strongly coupled. Indeed, concerning the zero-modes
the theory effectively behaves as a $D-1$ dimensional theory with effective coupling  ${g^2\over 2\pi
\e R_0}$ which blows up as $\e\to 0$. One should note that the treatment of the zero-modes in ref. \PH\ is
different from what people usually do in DLCQ, e.g. of QCD, which consists in first solving the
first-order equations of motion for the zero-modes and then plugging the solution back into the action,
generating instantaneous Coulomb-like interactions. Hellerman and Polchinski also note that in
certain supersymmetric QFTs the divergence of loop-diagrams when $\e\to 0$ can be avoided.
This raises the hope that M-theory might be well-behaved in this limit. They argue
that, if this limit exists, it should be the only reasonable way to define what one means by the DLCQ of
M-theory.

One might argue that M-theory  certainly is not an ordinary QFT and the analysis of Hellerman and
Polchinski need not be relevant in M-theory. This doubt is supported by the fact that M-theory
contains extended objects - membranes and five-branes - that can wrap around the compact
dimension. The existence of wrapping or winding modes is one of the crucial differences between
e.g. string and ordinary field
theory. Such winding states become very light (or of small effective tension in the $D-1$ dimensional
theory) as the compact circle shrinks. Said differently, as the radius goes to zero, more and more
winding modes contribute up to a given energy (tension) scale, and as $\e\to 0$ this may give rise to a
new divergence not present in {\it field} theory. Although we do not know how to precisely evaluate
this effect in M-theory, the presence of winding states is very familiar from string theory. 
Another crucial difference between field and string theory scattering amplitudes is the existence, in the
latter, of the moduli of the punctured Riemann surface which is the string world-sheet, 
that have to be integrated over. These integrations that translate the non-point-like character of string theory
tend to soften many of the field theory singularities.

So the best analogue for M-theory of the Hellerman-Polchinski one-loop amplitude probably
is some closed superstring
one-loop scattering amplitude with one spatial dimension compactified on a circle of radius $R=\e
\rap\equiv \e l_s$,
in the $\e\to 0$ limit\foot{
Up to now, we have called the radius of the space-like circle $R_s=\e R_0$, so the present choice
seems to imply  $R_0=l_s$.  Since $R_0$ does not have an invariant meaning this is of no relevance. Else
one can consider that we have rescaled $\e$, which also does not matter since in the end 
we are only interested in the $\e\to 0$ limit, anyhow.  
}. It is such an amplitude we will study in some detail below and show that the
$\e\to 0$ limit is finite and well-defined provided at least one external state has {\it non-vanishing}
momentum in the compact dimension. This is of course different from what one usually assumes in a
string compactification. However,  with view on the DLCQ of M-theory where the compact momenta are
$p_-^r=n_r/R_0$, this is just the case we are
interested in. Indeed, we showed above that the corresponding compact space-like momenta then are
$-p_1^r=n_r/(\e R_0)$ with the same (fixed) $n_r$.

\section{Summary}

Specifically, we will compute a four-point one-loop scattering amplitude in type II superstring theory
compactified on a spatial circle of small radius $R=\e \rap=\e l_s$. The external states are taken to have
arbitrary momenta $n_r/R$ in the compact direction, but no windings. The absence of windings of
the external states does not seem to be essential but simplifies the formulae.
With view on M-theory, we found it convenient to continue to call the mass of a state its
ten-dimensional mass\foot{
Our convention is such that $\mu$ runs over the nine non-compact dimensions 
$\mu= 0, 2, \ldots 8$, while $a\cdot b$ denotes a full ten-dimensional sum.
}, i.e. $M_r^2=-p_r\cdot p_r=-p_r^\mu p_{r\mu} -l_r^2/R^2$. One of the simplest computations 
then is the one with
all external states  being mass-less and having  factorized polarisations $\zeta^i_r
\overline\zeta^j_r$. This is the amplitude we will compute using the standard Green-Schwarz formalism. 
\foot{This standard Green-Schwarz light-cone formalism should of course not be confused with 
the light-cone we are
interested in here. In order to separate things as much as possible, note that 
the standard light-cone formalism
eliminates two sets of oscillators, say in the 0 and 9 directions, while the light-cone and light-like limit
we are interested in here concerns the 0 and 1 directions.
}
All conventions are as in Green-Schwarz-Witten [\GSW].

Our computation below shows some very general features and mechanisms that are clearly not
specific to a four-point amplitude, and we believe that they hold in a much more general context.
There are two competing effects as the radius of the circle shrinks to zero. 
The first is due to the winding modes running around the loop becoming
very light. This is the usual condensation of  light winding states that make divergent 
the naive compactification
of strings on a circle of vanishing size. Indeed there is one factor of $1/R=1/(\e l_s)$ just from
replacing the momentum integral by the discrete sum, and another from the condensation just
mentioned, alltogether giving a $1/\e^2$. One could still argue that after T-duality this is 
equivalent to an uncompactified theory, and
hence in terms of the T-dual coupling the result is finite. However, we want to keep the original
coupling constant fixed and then one  cannot escape the presence of an $1/\e^2$ factor.
The second effect, for non-vanishing external momenta $n_r/R$
in the compact direction, is the presence of 
a zero-mode factor $\exp\left[ -\pi Q(\nu_r,n,m)/\e^2\right]$ in the amplitude
where $Q$ is some positive definit complex quadratic form containing the
moduli $\nu_r$ of the 4-punctured torus, as well as the external momentum quantum numbers $n_r$ 
and the loop ``momentum and winding" quantum numbers $n$ and $m$. Generically, $Q\ne 0$ and as
$\e\to 0$ this exponential  vanishes. The main point is that, in combination with the $1/\e^2$ from
the first effect, this precisely combines to give  a complex $\delta^{(2)}(Q(\nu_r,n,m))$. The effect of this
$\delta$-function is to eliminate the  integration over one of the moduli, say $\nu_3$. However, one
still has the sum over $n$ and $m$, and the net effect will turn out to be that $\nu_3$ can now only
take finitely many discrete values on a lattice lying on the torus. So all that has happened in the
light-like limit is to discretise one of the moduli! We then check that this does not bring about any new
divergences and that the only singularities of the amplitude are those required by unitarity.
In particular, the discrete nature of $\n_3$ is just what is needed to produce the extra poles due to on-shell
intermediate states with non-vanishing winding numbers. We will also
show how the same amplitude can be obtained
by working with a light-like compactification from the outset,  after doing some suitable ``Wick rotation" 
of the otherwise divergent DLCQ amplitude.

So we conclude that in a setting relevant to DLCQ of M-theory (not all momenta in the compact
dimension vanishing) the one-loop scattering amplitude in type II superstring theory with four external
massless states has a finite and well-defined limit as the radius of the space-like compact  dimension
shrinks to zero. This limit coincides with the result obtained directly from a light-like compactification.
Since the mechanism just described seems to pertain not only to four-point amplitudes, we are
confident that it is a feature of any one-loop type II superstring amplitude, and probably also of all
higher genus ones as well. We take this as evidence that also in M-theory the light-like limit does exist and
coincides with its DLCQ. Of course, a more M-theoretic investigation is called for.

{\bf \chapter{The four-point one-loop amplitude for a space-like circle of vanishing radius}}

We will now describe the computation of the one-loop amplitude in some detail. We work within type II
superstring theory with one space-like dimension, say $x^1$, compactified on a circle of radius $R=\e l_s$ which we will
let go to zero in the end. The momenta of the four external states are denoted by $k_r$, $r=1,\ldots 4$, with
$\sum_{r=1}^4 k_r=0$, and
their polarisations are taken to factorize as $\zeta^{i_r}\overline \zeta^{j_r}$. The external states are
massless in the ten-dimensional sense, i.e $k_r\cdot k_r \equiv k_r^\mu k_{r,\mu} + n_r^2/R^2=0$ where
$n_r/R$ are the components of their momenta in the compact dimension. At least one of the $n_r$ (and hence
by momentum conservation actually at least two of them) are supposed to be non-vanishing. We suppose
that the {\it external} states have vanishing winding quantum numbers. It will be clear from the computation
below that non-vanishing winding numbers for the external states would not change the conclusion, but
would  slightly complicate the formulae. We will use the operator (Hamiltonian)
approach as extensively described in
[\GSW] to compute the amplitude, but we will eventually arrive at a form that could also be directly derived
from a path integral approach.

\sectionnumber=0
\section{The bosonic zero-modes}

The zero-mode part of the $L_0$ and $\overline L_0$ appearing in the string propagators are
$$L_0 = - {\ap\over 4}  p_0^2
+ {\ap\over 4}  \left( {n\over  R} - {m  R\over \ap}\right)^2 +{\ap\over 4} p_i^2 + \ {\rm oscillators}
\eqn\di$$
and similarly for $\overline L_0$, with $m\to -m$. This form of $L_0, \overline L_0$ corrseponds to the
Minkowski metric \i. We can rewrite them in a way that corresponds to the Lorenz-equivalent choice \iii\ by
letting  $p_0=\e p_t +{n\over R}$ (cf. \iv\ and remember $R=\e l_s$) so that 
$$L_0 = - {\ap\over 2} \left( {n\over l_s}+\e^2 p_t\right) \left( p_t +{m \over l_s}\right)
+ {\ap\over 4} \e^2 \left( p_t +{m \over l_s}\right)^2 +{\ap\over 4} p_i^2 + \ {\rm oscillators}
\eqn\dii$$
and again similarly for $\overline L_0$, with $m\to -m$. This latter form is more convenient when starting
directly with a compact light-like dimension because for $\e=0$ one simply gets
$L_0=-{1\over 2} n (l_s p_t+m) +{\ap\over 4} p_i^2 + \ {\rm oscillators}$. This will be used lateron. But first
we work with a space-like compactification, and hence with \di.

The amplitude contains a zero-mode piece
$$\prod_{r=1}^4 x_r^{{\ap\over 4} (p_r^{\rm R})^2} {\overline x_r}^{{\ap\over 4} (p_r^{\rm L})^2} \equiv F_1 F_2
\eqn\diii$$
where the  momenta in the loop are
$$\eqalign{
p_r^{{\rm R},\m} = p_r^{{\rm L},\m} \equiv p_r^\m &= p^\m -k_1^\m - \ldots - k_{r-1}^\m \cr
p_r^{{\rm R},1}&= {n\over R} - {R\over \ap}m -{n_1 + \ldots n_{r-1}\over R} \cr
p_r^{{\rm L},1}&= {n\over R} + {R\over \ap}m -{n_1 + \ldots n_{r-1}\over R}  \ .}
\eqn\div$$
The factors $F_1$ and $F_2$ respectively correspond to the contributions of the non-compact and
of the compact dimensions.
If we define as usual
$$\eqalign{ x_1 \ldots x_r = \r_r \ , \quad \r_4 \equiv w\ &, 
\quad w={\rm e}^{2\pi i \t}\ , \quad \r_r={\rm e}^{2\pi i \n_r} \cr
\nsr = \n_s-\n_r \ &, \quad \n_4\equiv \t  }
\eqn\dv$$
then $F_1$ and $F_2$ are given by
$$\eqalign{
F_1=&\exp\left\{ -\pi\ap\sum_{s>r} k_s^\m k_{r\m} \left[ {(\Im \nsr )^2\over \Im \t } - \Im \nsr \right] \right\} 
\exp\left\{ -\pi\ap \Im\t \left( p^\m +\sum_s k_s^\m {\Im\n_s\over \Im\t} \right)^2 \right\}\cr
F_2=&\exp\left\{ {i\pi\over 2} \ap \sum_{s>r} {n_s n_r \over R^2} 
\left( {\nsr^2\over \t} - \nsr - {\overline\nsr^2\over \tb} + \overline \nsr \right) \right\} \cr
& \times 
\exp\left\{ {i\pi\over 2} \t \ap \left( {n\over R}-{R m\over  \ap} + \sum_s {n_s \n_s\over R \t} \right)^2
- {i\pi\over 2} \tb \ap \left( {n\over R}+{R m\over  \ap} + \sum_s {n_s \overline\n_s\over R \tb} \right)^2
\right\} \ . \cr }
\eqn\dvi$$
(Sums over $s$ or $r$ always run from 1 to 4.)
Note that in the usual string theory compactification with all $n_r$ vanishing the first exponential
factor in $F_2$ is absent. 
One needs to compute $\left( \int {\rm d}^9 p F_1 \right) \left( {1\over R} \sum_{m,n} F_2 \right)$. In
the integral involving $F_1$ one simply shifts the integration variable to obtain as usual
$$ \int {\rm d}^9 p\,  F_1 = (\ap \Im\t)^{-9/2} \prod_{s>r} 
\left[ \exp\left\{ -\pi \left[ {(\Im \nsr )^2\over \Im \t } - \Im \nsr  \right] \right\} \right]^{\ap k_s^\m k_{r\m}} \ .
\eqn\dvii$$
The sum over $n,m$ of $F_2$ is more similar to the lattice sum for the heterotic string, except that here we
have a $\t,\n$ and a $\tb,\overline\n$ part.

\section{The amplitude}

The contributions of the fermionic zero-modes and of the non-zero modes, bosonic and fermionic, are the
standard ones, see [\GSW]. They give rise to the kinematic factor
$$\eqalign{
\kcl&=K({k_r\over 2}, \zeta_r) K({k_r\over 2}, \overline\zeta_r) \cr
K({k_r\over 2}, \zeta_r)&=\zeta_1^{i_1} \ldots \zeta_4^{i_4} \tr ( R_0^{i_1 j_1}\ldots R_0^{i_4  j_4})
{k_1^{j_1}\over 2} \ldots {k_4^{j_4}\over 2} }
\eqn\dviii$$
with the $R_0^{ij}$ defined in [\GSW], and the factors
$$ \prod_{s>r} \chi(\nsr,\t)^{\ap k_r\cdot k_s} 
\eqn\dix$$
where $\chi(\n,\t)$ is defined by
$$\chi(\n,\t)=2\pi \exp\left\{ -\pi {(\Im\n)^2\over \Im\t} \right\} 
\left\vert {\theta_1(\n,\t) \over\theta_1'(\n,\t)} \right\vert \ .
\eqn\dx$$
In \dix\ each single factor  includes a piece 
$\exp\left\{ -\pi \left[ {(\Im \nsr )^2\over \Im \t } - \Im \nsr  \right] \right\}^{\ap n_s n_r /R^2}$
which should have come from \dvii\ if we had no compact dimension, or else which would be equal to one if
the $n_r$ would vanish. Here we have included this piece by hand in order that full ten-dimensional sclar
products $k_r\cdot k_s$ appear in \dix, so we have to divide this piece out again. Putting everything together,
we obtain for the full four-point one-loop amplitude
$$\eqalign{
A_{\rm cl}^{(4)}&=(\pi \kappa)^4 \kcl \int {\rm d}^2\t\, {\rm d}^2\n_1\, {\rm d}^2\n_2\, {\rm d}^2\n_3 \ I \cr
I&=(\ap \Im\t)^{-9/2}  \prod_{s>r} \chi(\nsr,\t)^{\ap k_r\cdot k_s}\ J \cr
J&=\exp\left\{ \pi \ap \sum_{s>r} {n_s n_r \over R^2} 
\left[ {(\Im \nsr)^2\over \Im\t } -{\nsr^2\over 2i\t} + {\overline\nsr^2\over 2i\tb} \right] \right\}\  {\cal S} \cr
{\cal S}&={1\over R} \sum_{n,m} 
\exp\left\{ {i\pi\t\over 2}  \ap \left( {n\over R}-{R m\over  \ap} + \sum_s {n_s \n_s\over R \t} \right)^2
- {i\pi\tb\over 2}  \ap \left( {n\over R}+{R m\over  \ap} + \sum_s {n_s \overline\n_s\over R \tb} \right)^2
\right\} \ .\cr }
\eqn\dxi$$
It will be useful to rewrite $J$, using the identity $\sum_{s>r} n_r n_s \nsr^2 ={1\over 2} \sum_{r,s} n_r n_s
\nsr^2=-(\sum_r n_r \n_r)^2$ (since $\sum n_r=0$), as
$$\eqalign{
J&=\exp\left\{ -\pi \ap \sum_{s,r} {n_s n_r \over R^2} 
\left[ {\Im \n_s \Im \n_r\over \Im\t } - \Im{\n_s\n_r\over \t} \right] \right\}  {\cal S}\cr
&=\exp\left\{ -\pi \ap \sum_{s,r} {n_s n_r \over R^2}  
{\Im (\n_s/\t) \Im (\n_r/\t)\over \Im (-1/\t) }\right\}  {\cal S} \ .\cr }
\eqn\dxii$$

\section{Modular invariance}

It is easy to verify that the new factors involving the $n_r$ do not spoil modular invariance, thus providing a
check of the above expression for the amplitude. First, invariance under $\n_r\to \n_r +1$ and under
$\n_r\to \n_r +\t$ follows trivially from the standard properties of the $\chi(\n,\t)$. Invariance under $\t\to \t+1$
no longer is manifest, but was evident initially in \diii\ because
${\ap\over 4} (p_r^{\rm R})^2 -{\ap\over 4} (p_r^{\rm L})^2$ is an integer. To check invariance under $\t\to
-1/\t,\ \n_r\to -\n_r/\t$ one has to perform a standard Poisson resummation of ${\cal S}$:
$${\cal S}\left(-{\n_r\over \t}, -{1\over \t}\right)= \vert \t \vert 
\exp\left\{\pi \ap \sum_{s,r} {n_s n_r \over R^2}  \Im{\n_s\n_r\over \t}  \right\}  {\cal S}(\n_r,\t)
\eqn\dxiii$$
so that using the form \dxii\ of $J$ it is obvious that $J(-\n_r/\t,-1/\t)=\vert\t\vert J(\n_r,\t)$ and thus
$I(-\n_r/\t,-1/\t)=\vert\t\vert^{10} I(\n_r,\t)$ which proves modular invariance of the amplitude.
Consequently, as usual, $\t$ is to be integrated over the standard fundamental domain, and each of the $\n_r$
over the parallelogram with corners $(0,1,1+\t,\t)$. In other words, for each given $\t$ determining the shape
of the world-sheet torus, the $\n_1, \n_2, \n_3$ are to be integrated over this torus (with $\n_4$ being fixed
at $\t$ or equivalently at 0).

\section{Path integral form of the amplitude}

We now rewrite the sum ${\cal S}$ by performing  a partial Poisson resummation in $m$ only. This yields
$$\eqalign{
{\cal S}&=\left( {\ap\over \Im\t}\right)^{1/2}\,  {1\over R^2} \exp\left\{ \pi \ap \sum_{s,r} {n_s n_r \over R^2}  
{\Im (\n_s/\t) \Im (n_r/\t)\over \Im (-1/\t) }\right\}\cr
&\times \sum_{n,m} \exp\left\{ -\pi {\ap \over R^2}\left[ \Im\t \left( n+ \sum_s n_s {\Im\n_s\over \Im\t}\right)^2
+{1\over \Im\t} \left( m+\Re\t n +\sum_s n_s \Re\n_s\right)^2 \right]\right\} \ . \cr}
\eqn\dxiv$$
The first exponential cancels against the one in $J$ and the second exponential can be rearranged so that
$$J=(\ap \Im\t)^{-1/2}\,  {\ap\over R^2} \sum_{n,m} 
\exp\left\{ -\pi {\ap \over R^2} {1\over \Im\t} \left\vert m+n\t +\sum_s n_s\n_s \right\vert^2\right\}
\eqn\dxv$$
which is now quite simple and which is the form one would have gotten directly from a path-integral
computation. Also, the modular properties of $J$ now are manifest. Inserting this form of $J$ into eq. \dxi\
we finally get for the amplitude
$$\eqalign{
A_{\rm cl}^{(4)}=&{(\pi \kappa)^4 \over \ap^5} \kcl 
\int {{\rm d}^2\t\over (\Im\t)^2}\,  \prod_{r=1}^3 {{\rm d}^2\n_r\over \Im\t}\
\prod_{s>r} \chi(\nsr,\t)^{\ap k_r\cdot k_s}  \cr
&\times \sum_{n,m} 
{\ap\over R^2} \exp\left\{ -\pi {\ap \over R^2}
{1\over \Im\t} \left\vert m+n\t +\sum_s n_s\n_s \right\vert^2 \right\} \cr}
\eqn\dxvi$$

\section{The limit of the space-like circle of zero radius}

So far, the radius $R$ of the compact space-like dimension was arbitrary. Now we will study what happens
if we let 
$$R^2/\ap=\e^2\to 0 \ .
\eqn\dxvii$$
The form \dxvi\ of the amplitude is particularly convenient to study this limit. 
First, let us make some preliminary
remarks. Obviously, we have to study the $\e\to 0$ limit of
$${1\over \e^2}  \exp\left\{ -{\pi\over \e^2}  {1\over \Im\t} 
\left\vert m+n\t +\sum_s n_s\n_s \right\vert^2 \right\} \ .
\eqn\dxviii$$
The $1/\e^2$ in the exponential  ensures that one can get a non-vanishing contribution only if 
$m+n\t +\sum_s n_s\n_s=0$. If all $n_s$ vanish (the usual case studied in string compactification on a circle)
then only $m=n=0$ can contribute to the sum, in which case the exponential in \dxviii\  just gives 1 and we are
left with the $1/\e^2$ factor leading to the well-known divergence discussed in the introduction.
For non-vanishing $n_s$ however, the argument of the exponential depends on the integration variables
$\n_r$ and things are more subtle.

In fact, the factors of $\e$ are precisely such that in the $\e\to 0$ limit one obtains a delta-function:
$${1\over \e^2}  \exp\left\{ -{\pi\over \e^2}  {1\over \Im\t} 
\left\vert m+n\t +\sum_s n_s\n_s \right\vert^2 \right\} 
\ \to \  \Im\t\,  \delta^{(2)} \left(  m+n\t +\sum_s n_s\n_s \right) \ ,
\eqn\dxix$$
so that we arrive at
$$A_{\rm cl}^{(4)}\Big\vert_{\e\to 0}={(\pi \kappa)^4 \over \ap^5} \kcl 
\int {{\rm d}^2\t\over (\Im\t)^2}  \prod_{r=1}^3 {{\rm d}^2\n_r\over \Im\t}\,
\prod_{s>r} \chi(\nsr,\t)^{\ap k_r\cdot k_s}\ \Im\t
\sum_{m,n} \delta^{(2)} \left(  m+n\t +\sum_s n_s\n_s \right) 
\eqn\dxixa$$
The delta-function  suppresses one full complex integration over one modulus $\n_r$ provided not all $n_r$
vanish. If all $n_r$ vanish, one just gets $\delta^{(2)} \left(  m+n\t \right)$ singling out $m=n=0$ and giving a
divergent $\delta^{(2)}(0)\sim 1/\e^2$ as before. However, as already stressed, we are interested in the case where at
least some $n_r\ne 0$. To be concrete, let's assume $n_3\ne 0$, otherwise relabel the external states.
We then want to trade the delta-function against the $\n_3$-integration. 
While the the form \dxixa\ of the amplitude still was manifestly symmetric under exchange of the external
particles, this will of course no longer be true in the following. In the sum, only those $m,n$ can
contribute that are such that $\n_3$ is within its integration region, namely the parallelogram $(0,1,1+\t,\t)$.
Since 
$$ m+n\t +\sum_s n_s\n_s = n_3 \left( \n_3-{\n_{41}n_1+\n_{42}n_2-m-n\t\over n_3} -\t \right)
\eqn\dxx$$
there are precisely $n_3$ values of $m$ and $n_3$ values of $n$ that contribute, thus $n_3^2$ discrete
values of $\n_3$ that contribute to the sum. These $n_3^2$ values of $\n_3$ fill out a finite regular lattice
within the parallelogram $(0,1,1+\t,\t)$. If we denote by
$\tilde \n_0$ the point among them that is closest to the origin then for any function $f$ of $\n_3$
$$\int {\rm d}^2 \n_3 \sum_{m,n=-\infty}^\infty \delta^{(2)} \left(  m+n\t +\sum_s n_s\n_s \right)  f(\n_3)
={1\over n_3^2} \sum_{m,n=0}^{n_3-1} f\left(\tilde \n_0+{m+n\t\over n_3}\right) \ .
\eqn\dxxi$$
Of course, $\tilde \n_0$ depends on $\n_1, \n_2, \t$ as well as on the $n_r$. A convenient way to characterise
$\tilde \n_0$ is the following: denote by $F[x]=x-E[x]$ the
fractional part of the real number $x$, and for every complex number of the form $z=x+\t y$ with real $x,y$,
let $F_c[z]=F[x] + \t F[y]$. Then one simply has
$$\tilde \n_0 = {1\over n_3} F_c[\n_{41} n_1 + \n_{42} n_2]\ .
\eqn\dxxii$$

Putting everything together we find that the $\e\to 0$ limit of the amplitude is
$$
A_{\rm cl}^{(4)}\Big\vert_{\e\to 0}={(\pi \kappa)^4 \over \ap^5} \kcl 
\int {{\rm d}^2\t\over (\Im\t)^2}\, { {\rm d}^2\n_1\over \Im\t}\,  {{\rm d}^2\n_2\over \Im\t}\,
{1\over n_3^2} \sum_{m,n=0}^{n_3-1} 
\prod_{s>r} \chi(\nsr,\t)^{\ap k_r\cdot k_s} \Big\vert_{\n_3=\tilde \n_0 +{m+n\t\over n_3}} \ .
\eqn\dxxiii$$
We see that the only effect of the compactification on a space-like circle  of vanishing size (with external momenta in the
compact direction being $n_r/(\e l_s)$, $n_r$ being kept fixed) is to simply replace one of the $\n_r$
integrations by a discrete sum over $n_r^2$ values on a regular lattice on the torus, i.e the parallelogram
$(0,1,1+\t,\t)$.  This is quite striking. In particular, the amplitude \dxxiii\ is perfectly finite. Of course one still
has to check that the integrations over the remaining moduli do not induce any new divergences.  It is
however easy to verify that the only singularities of the amplitude \dxxiii\ are those poles that are compatible
with unitarity, corresponding to on-shell intermediate states. This will be discussed next.

\section{Finiteness of the amplitude in the light-like limit}

Possible divergences of the amplitude \dxxiii\ arise whenever two or more of the $\n_r$ come close to each
other, which now in particular also means $\n_1$ or $\n_2$ close to any of the discrete values of $\n_3$, or 
$\n_1$ and $\n_2$ such that $\n_3$ is close to $\t\equiv \n_4$. All these singularities can be easily studied
using the asymptotic form of 
$$\chi(\nsr,\t) \sim 2\pi \vert \nsr \vert\qquad {\rm as } \ \nsr \to 0 \  .
\eqn\dxxiv$$
We have checked that the only divergences that arise are the poles required by unitarity. Here we will only
present one example which corresponds to the case studied in field theory in [\PH]. This is the case were
there is vanishing momentum transfer in the compact direction between the two scattering particle. Here this
corresponds to the four-point amplitude with e.g. $n_1=-n_2=\tilde l$ (first particle) and $n_3=-n_4=l$
(second particle). The dangerous field theory diagram then corresponds to the limit where $\n_{21}\to 0$. Of
course, the $\n_{21}=0$ limit of the string amplitude integrand is divergent, but what we must do is to carry
out the integral and check whether the small $\n_{21}$ region gives a divergence or not. 

Note that with the present choice of $n_r$ one has $(\n_{41} n_1+\n_{42}n_2)= \tilde l \n_{21}$ so that
in the region of interest where $\n_{21}$ is small, this quantity is small as well. Then
$$\tilde\n_0={1\over l}F_c[\tilde l\, \n_{21}]={\tilde l\over l}\, \n_{21}\quad {\rm and} \quad
\n_3={\tilde l\,  \n_{21} +m+m\t\over l} \ .
\eqn\dxxv$$
First, for $m$ and $n$ not both zero, $\n_3$ will not be close to any other $\n_r$ in general, and the discrete nature
of $\n_3$ plays no special role, so that one only gets divergences from
$$\int {\rm d}^2\n_{21}\,  \chi(\n_{21},\t)^{\ap k_1\cdot k_2} \sim 
\int {\rm d}^2\n_{21}\,  (2\pi \vert  \n_{21} \vert ) ^{\ap k_1\cdot k_2} \sim {1\over 2- \ap t /2}
\eqn\dxxvi$$
where we introduced the Mandelstam variable 
$$t=-(k_1+k_2)^2=-2 k_1\cdot k_2 =-2 k_3\cdot k_4\ .
\eqn\dxxvii$$
Thus there are poles for states in the $t$-channel that have mass squared equal $4/\ap$. With the
conventions used [\GSW] for the closed string, this is just the first massive level of the uncompactified
closed superstring. Thus this pole must well be
there for unitarity.

However, the compactified closed superstring has more massive levels. We always refer to the
ten-dimensional mass. The point is that in the limit we consider, the left-right level matching condition no
longer is $N_{\rm oscill}=\overline N_{\rm oscill}$ thus forcing the total  level to be even, but rather
 $N_{\rm oscill}- nm/2 = \overline N_{\rm oscill} + nm/2$ thus allowing any integer level number.
In particular we expect the first massive pole to appear already at mass squared equal to $2/\ap$, i.e. at 
$\ap t=2$ rather than $\ap t =4$. We will now show that it is precisely the discrete nature of the moduli $\n_3$
that gives rise to these new poles, and hence remembers that one dimension was compactified so that there
were winding states running around the loop!

Above, we have looked at $m$ and $n$ not both zero. Now consider $n=m=0$. Then $\n_3={\tilde l\over
l}\n_{21}$. So as long as $\n_{21}$ is very small $\n_3$ is also forced to be very small, meaning it is
very close to zero, which by the periodicity of the $\chi$ is equivalent to being very close to $\t$. Hence also
$\n_{43}$ is very small. Rather than being an independent integration variable, $\n_{43}$ is driven to zero if
$\n_{21}$ is taken to zero. This changes the nature of the singularity to be
$$\eqalign{&\int {\rm d}^2\n_{21}\,  \chi(\n_{21},\t)^{\ap k_1\cdot k_2} \chi(\n_{43},\t)^{\ap k_3\cdot k_4} \cr
&\sim \int {\rm d}^2\n_{21}\,  (2\pi \vert  \n_{21} \vert ) ^{-\ap t/2}
(2\pi (\tilde l/l) \vert  \n_{21} \vert ) ^{-\ap t/2}  \sim {1\over 2- \ap t } \ . }
\eqn\dxxviii$$
Thus we indeed see the desired pole at $\ap t=2$.

{\bf \chapter{Direct DLCQ computation of the four-point one-loop amplitude}}

In this section we will show how to compute the four-point one-loop amplitude directly with a compactified
light-like circle. In the first place, the result will be very singular, however. The reason is that it is not clear
how to naturally implement the Wick rotation that was implicitly made above when we computed $\int {\rm
d}p_0 F_1$. Very formally, we can then just replace the integration over the continuous light-cone momentum
by a ``Wick-rotated" one ($p\to i p$). The resulting expression can then be shown to coincide with the above
amplitude $A_{\rm cl}^{(4)}\vert_{\e\to 0}$ of eq. \dxxiii. The fact that both expressions coincide can be taken
as an {\it a posteriori} justification of this ``Wick rotation". In this sense we can say that in type II string
theory the DLCQ can indeed be viewed as the limit of an almost light-like compactification.

As extensively discussed in the introduction, doing a DLCQ computation amounts to taking $\e=0$ from the
outset, in particular before doing the zero-mode integration $\int {\rm d}p_0$. We will here simply use the
formalism of the preceeding section and examine how to set $\e=0$ from the start. The reader should be
warned that most of this is very formal, since e.g. we used the representation of the $r^{\rm th}$
string propagator ${1\over L_0}$ as $\int_0^1 {\rm d}x_r x_r^{L_0-1}$ which can be justified only after a
Wick rotation of $p_0$ that makes $L_0$ positive. On the other hand, in DLCQ, $L_0$ is indefinite, and
hence the whole procedure remains formal. Nevertheless it is interesting to see that in the end, after these
formal manipulation one obtains exactly the same amplitude as was derived quite rigorously in the
preceeding section.

As noted in the beginning of section 2, the DLCQ form of $L_0$ is obtained from \dii\ with $\e=0$. Therefore
one needed to change the momentum variable from $p_0$ to $p_t$ by $p_0=\e p_t +{n\over \e l_s}$. The
resulting $L_0$ and $\overline L_0$ are
$$\eqalign{
L_0&=-{1\over 2} n (l_s p_t+m) +{\ap\over 4} p_i^2 + \ {\rm oscillators}\cr
\overline L_0&=-{1\over 2} n (l_s p_t-m) +{\ap\over 4} p_i^2 + \ {\rm oscillators} }
\eqn\ti$$
The bosonic zero-mode piece $F_1F_2$ coming from \diii\ then is modified. 
In practice, the easiest way to obtain it is to rewrite \dvi\ as 
$$\eqalign{
F_1 F_2 &= \exp\left\{ -\pi\ap\sum_{s>r} k_s\cdot k_r \left[ {(\Im \nsr )^2\over \Im \t } - \Im \nsr \right] \right\} 
\exp\left\{ -2\pi i m \Re \left( n\t +\sum_s n_s \n_s \right)\right\}  \cr
&\times
\exp\left\{ -\pi \Im\t \left[ \ap \left( p^\m +\sum_s k_s^\m {\Im\n_s\over \Im\t}\right)^2
+ {1\over \e^2}\left( n+ \sum_s n_s {\Im\n_s\over \Im\t}\right)^2  +\e^2 m^2 \right] \right\}   }
\eqn\tii$$
and to change the loop momentum variable from $p_0$ to
$$\e p_t= p_0 -{n\over \e l_s} + \sum_s \left( k_s^0 -{n_s\over \e l_s} \right) {\Im\n_s\over \Im\t}
\eqn\tiii$$
so that one can now safely set $\e=0$ and obtain
$$\eqalign{
F_1 F_2 \Big\vert_{\e=0}
&= \exp\left\{ -\pi\ap\sum_{s>r} k_s\cdot k_r \left[ {(\Im \nsr )^2\over \Im \t } - \Im \nsr \right] \right\}  \cr
&\times \exp\left\{ 2\pi\,  l_s\,  p_t\,  \Im \left( n\t +\sum_s n_s \n_s \right) 
-2\pi i\,  m\,  \Re \left( n\t +\sum_s n_s \n_s \right)  \right\}  \cr
&\times \exp\left\{ -\pi \ap \Im\t  \left( p^j +\sum_s k_s^j {\Im\n_s\over \Im\t}\right)^2 \right\} }
\eqn\tiv$$
where the indices $j$ only run over the eight transverse dimensions, $j=2, \ldots 9$.
On the other hand, the integration and sum are replaced by
$${\rm d}p_0\, {1\over \e l_s}\,  \sum_{n,m} = {1\over l_s}\,  {\rm d} p_t\,  \sum_{n,m} \ ,
\eqn\tv$$
which  also is independent of $\e$.

Adding as before the contributions \dix\ of the non-zero modes, as well as of the fermionic zero-modes 
\dviii, the amplitude in the DLCQ reads (we write $p= l_s p_t$)
$$\eqalign{
A_{\rm cl, \, {\rm DLCQ}}^{(4)}
&={(\pi \kappa)^4\over \ap^5} \kcl 
\int {\rm d}^2\t\, {\rm d}^2\n_1\, {\rm d}^2\n_2\, {\rm d}^2\n_3 \, (\Im\t)^{-4} 
\prod_{s>r} \chi(\nsr,\t)^{\ap k_r\cdot k_s} S_{\rm DLCQ} \cr
S_{\rm DLCQ}&=\int {\rm d} p \sum_{n,m} \exp\left\{ 2\pi p\, \Im \left( n\t +\sum_s n_s \n_s \right) 
-2\pi i m\, \Re \left( n\t +\sum_s n_s \n_s \right)  \right\} . }
\eqn\tvi$$
The quantity $S_{\rm DLCQ}$ looks awfully divergent due to the absence of a Wick rotation as discussed
above. Comparing with the amplitude  \dxixa\ computed in the previous section, we see
that we want to interpret  $S_{\rm DLCQ}$ as $\sum_{m,n}\delta^{(2)}\left(m+n\t +\sum_s n_s \n_s \right)$.
But this can easily be achieved. All one has to do is to ``Wick rotate" the light-cone $p$ as $p\to i p$. Then
$$S_{\rm DLCQ} \to \sum_n \int {\rm d} p\, \exp\left\{ 2\pi i\,  p\,  \Im  \left( n\t +\sum_s n_s \n_s \right) \right\}
\sum_m  \exp\left\{ -2\pi i\,  m\,  \Re  \left( n\t +\sum_s n_s \n_s \right) \right\} 
\eqn\tvii$$
While the integral over $p$ of the first exponential just gives 
$\delta \left(  \Im  \left( n\t +\sum_s n_s \n_s \right) \right)$,
the sum over $m$ of the  second exponential gives a periodic delta-function
so that
$$\eqalign{
S_{\rm DLCQ}^{\rm ``Wick"}&= \sum_n
\delta \left(  \Im \left(  n\t +\sum_s n_s \n_s \right) \right)
\sum_{\tilde m}
\delta \left(  \tilde m + \Re \left(  n\t +\sum_s n_s \n_s \right) \right)\cr
&=\sum_{n,\tilde m} \delta^{(2)}\left(\tilde m+n\t +\sum_s n_s \n_s \right) \ .  }
\eqn\tviii$$
The resulting four-point one-loop amplitude then is identical with the one derived in the previous section in
the light-like {\it limit}, $\e\to 0$, i.e. with \dxixa\  or \dxxiii.

\ack
I am grateful to J.L.F. Barbon for extensive and fruitful  discussions, as well as for useful criticism on
an earlier version of this manuscript. I also profited from
conversations with K. Sfetsos and L. Susskind. I would like to thank the CERN Theory group for
hospitality and for organizing a very stimulating workshop. This work was partially supported by the
European Commision under TMR contract FMRX-CT96-0090.

\refout

\end